\documentclass[letterpaper]{article}

\usepackage{fancyhdr}
\usepackage[letterpaper,margin=1.0in,headheight=0.2in, includefoot]{geometry}
\usepackage{titling}
\usepackage{enumitem}
\usepackage{amsmath}
\usepackage{graphicx}
\usepackage{hyperref}
\usepackage{cleveref}
\usepackage{caption}
\usepackage{authblk}
\usepackage{abstract}
\usepackage{appendix}
\usepackage{acro}
\DeclareAcronym{nasa}{
	short = NASA,
	long = National Aeronautics and Space Administration,
}

\DeclareAcronym{esa}{
	short = ESA,
	long = European Space Agency,
}

\DeclareAcronym{mule}{
	short = MULE,
	long = Microreactor Utilisation for Lunar Exploration,
}

\DeclareAcronym{eisi}{
	short = EISI,
	long = ESA Initial Support for Innovation ,
}

\DeclareAcronym{isru}{
	short = ISRU,
	long = in-situ resource utilisation ,
}

\DeclareAcronym{trl}{
	short = TRL,
	long = technology readiness level,
}

\DeclareAcronym{mse}{
	short = MSE,
	long = molten salt electrolysis,
}

\DeclareAcronym{mre}{
	short = MRE,
	long = molten regolith electrolysis,
}

\DeclareAcronym{tum}{
	short = TUM,
	long = {Technical University of Munich},
}

\DeclareAcronym{cfd}{
	short = CFD,
	long = computational fluid dynamics,
}

\DeclareAcronym{triso}{
	short = TRISO,
	long = tristructural-isotropic,
}

\DeclareAcronym{sic}{
	short = SiC,
	long = silicon carbide,
}

\DeclareAcronym{beo}{
	short = BeO,
	long = beryllium oxide,
}

\DeclareAcronym{uc}{
	short = UC,
	long = uranium carbide,
}

\DeclareAcronym{heu}{
	short = HEU,
	long = highly enriched uranium,
}

\DeclareAcronym{b4c}{
	short = B$_4$C,
	long = boron carbide,
}

\DeclareAcronym{cad}{
	short = CAD,
	long = computer-aided design,
}

\DeclareAcronym{bol}{
	short = BOL,
	long = begin of life,
}

\DeclareAcronym{eol}{
	short = EOL,
	long = end of life,
}

\usepackage[separate-uncertainty=true]{siunitx}
\usepackage[T1]{fontenc}
\usepackage{verbatim}
\usepackage{subcaption}
\usepackage{placeins}
\usepackage{ifthen}

\newcommand{\keff}{\ensuremath{k_{\mathrm{eff}}}}
\newcommand{\uncpcm}[2]{\num{#1}$\ \pm\ $\SI{#2}{pcm}}

\begin{document}

\title{\textbf{\acs{mule} --- A Co-Generation Fission Power Plant Concept to Support Lunar In-Situ Resource Utilisation}}

\author[1]{Julius Mercz}
\author[2]{Philipp Reiss}
\author[1]{Christian Reiter}

\affil[1]{Professorship of Applied Nuclear Technologies, \acf{tum}, \hbox{Boltzmannstr. 15}, 85748 Garching, Germany}
\affil[2]{Professorship of Lunar and Planetary Exploration, \acf{tum}, \hbox{Lise-Meitner-Str. 9}, 85521 Ottobrunn, Germany}

\predate{}
\postdate{}
\date{}

\fancyhf{}
\thispagestyle{empty}
\pagestyle{empty}

\fancypagestyle{firstpage}{
	\fancyhead[C]{
		
	}
	\lfoot{}
	\rfoot{}
}
\thispagestyle{firstpage}

\pagestyle{fancy}

\fancyhead[C]{\ifthenelse{\isodd{\value{page}}}{\small \acs{mule} --- A Co-Generation Fission Power Plant Concept to Support Lunar In-Situ Resource Utilisation}{\small J. Mercz, P. Reiss \& C. Reiter}}

\rfoot{}
\cfoot{\thepage}
\lfoot{
	
}

\renewcommand{\headrulewidth}{0pt}  
\urlstyle{rm}
\setlength{\parskip}{5pt}   
\setlength\parindent{0pt} 
\setlist{nolistsep}
\renewcommand{\thetable}{\Roman{table}}

\maketitle

\begin{abstract}
For a sustained human presence on the Moon, robust \acl{isru} supply chains to provide consumables and propellant are necessary.
A promising process is \acl{mse}, which typically requires temperatures in excess of \num{900}\ °\si{\celsius}.
Fission reactors do not depend on solar irradiance and are thus well suited for power generation on the Moon, especially during the 14-day lunar night.
As of now, fission reactors have only been considered for electric power generation, but the reactor coolant could also be used directly to heat those processes to their required temperatures.
In this work, a concept for a co-generation fission power plant on the Moon that can directly heat a \ac{mse} plant to the required temperatures and provide a surplus of electrical energy for the lunar base is presented.
The neutron transport code \textit{Serpent 2} is used to model a ceramic core, gas-cooled very-high-temperature microreactor design and estimate its lifetime with a burnup simulation in hot conditions with an integrated step-wise criticality search.
Calculations show a neutronically feasible operation time of at least 10 years at \SI{100}{\kilo\watt} thermal power.
The obtained power distributions lay a basis for further thermal-hydraulic studies on the technical feasibility of the reactor design and the power plant.
\end{abstract}
\providecommand{\keywords}[1]
{
  \small	
  \textbf{\textit{Keywords---}} #1
}

\section{Introduction} \label{sec:intro}

In the upcoming years, human lunar surface activity is likely to increase again, with programmes like \textit{Artemis} from the \ac{nasa}, \textit{Terrae Novae strategy} from the \ac{esa} or the multinational \textit{International Lunar Research Station}.
These planned missions will increase the demand of power on the Moon by orders of magnitude.
In particular, long-term crewed bases require a continuous and reliable power source, to ensure astronaut safety and enable unrestricted operation of all necessary life support systems even during the 14-day lunar night.
To advance independence of sustained human presence on other planetary bodies, \acf{isru} techniques will be employed to provide consumables and propellant \cite{Sacksteder2007}.
These are extracted from the local regolith, which is the uppermost layer of a planetary surface consisting of "unconsolidated, weathered, broken rock debris, mineral grains, and superficial deposits which overlie the unaltered bedrock" \cite{Mayhew2009}.
Such processes often require high temperatures in excess of \SI{900}{\celsius} \cite{Chen2000, Sargeant2020}. 

The power requirement can be met by nuclear fission reactors, which can supply large amounts of electricity and heat on a constant level for a long period of time, while being compact in size \cite{IAEA2020}.
Its independence from an external energy source, like for example solar irradiance, increase mission robustness and reduces the necessary auxiliary electrical storage capacity.
It provides more flexibility in mission planning by extending the spectrum of possible landing sites to less illuminated polar regions, permanently shadowed craters or cave structures.
Furthermore, it enables long-term missions with increased power demands to the outer solar system, where solar irradiance is decreasing.

As of now, nuclear fission reactors in spaceflight have only been utilised in satellites between the 1960s to 1980s for providing electric power via thermo-electric generators with low efficiencies, providing only a few \si{\kilo\watt_e} \cite{Staub1973}.
In the following decades, solar panels, and infrequently radioisotope generators, were favoured over fission reactors due to advancements in solar cell technology, as well as the lack of spaceflight missions with high power requirements.
During this period, fission reactors for space missions were studied only conceptually.
The returned interest in crewed missions to the Moon and Mars consolidated the investigation of nuclear power systems, with the notable example of the US project \textit{Kilopower}, now \textit{Fission Surface Power}, which demonstrated a \acl{trl} of 5 during the successful experimental \textit{KRUSTY} campaign \cite{Gibson2018}.
These projects aim at developing fission power systems with up to \SI{40}{\kilo\watt_{e}} and a lifetime of 10 years, using Stirling power conversion systems \cite{Gibson2014, NASA2022}.

Besides generating electric power, fission power systems have not been considered to supply process heat directly.
In particular, heating of medium- to high-temperature \ac{isru} processes like the reduction of regolith using hydrogen or methane, \acf{mse}, or even \ac{mre} via a fission power system has not yet been investigated.
An additional benefit from such a high-temperature reactor is the ability to use the remaining enthalpy of its coolant in a subsequent electric conversion process and heating of base habitats and facilities or potential thermal energy storages that provide additional redundancy.
To mature this concept, \ac{tum} has started the \acf{mule} project to design a fission power system for this purpose, as well as the necessary simulation framework \cite{Mercz2024}.

\FloatBarrier 
\section{MULE Project \& Core Design} \label{sec:mule}

\subsection{Project Overview} \label{ssec:mule_overview}

\begin{figure}[ht]
	\center{
		\includegraphics[width=0.93\textwidth]{"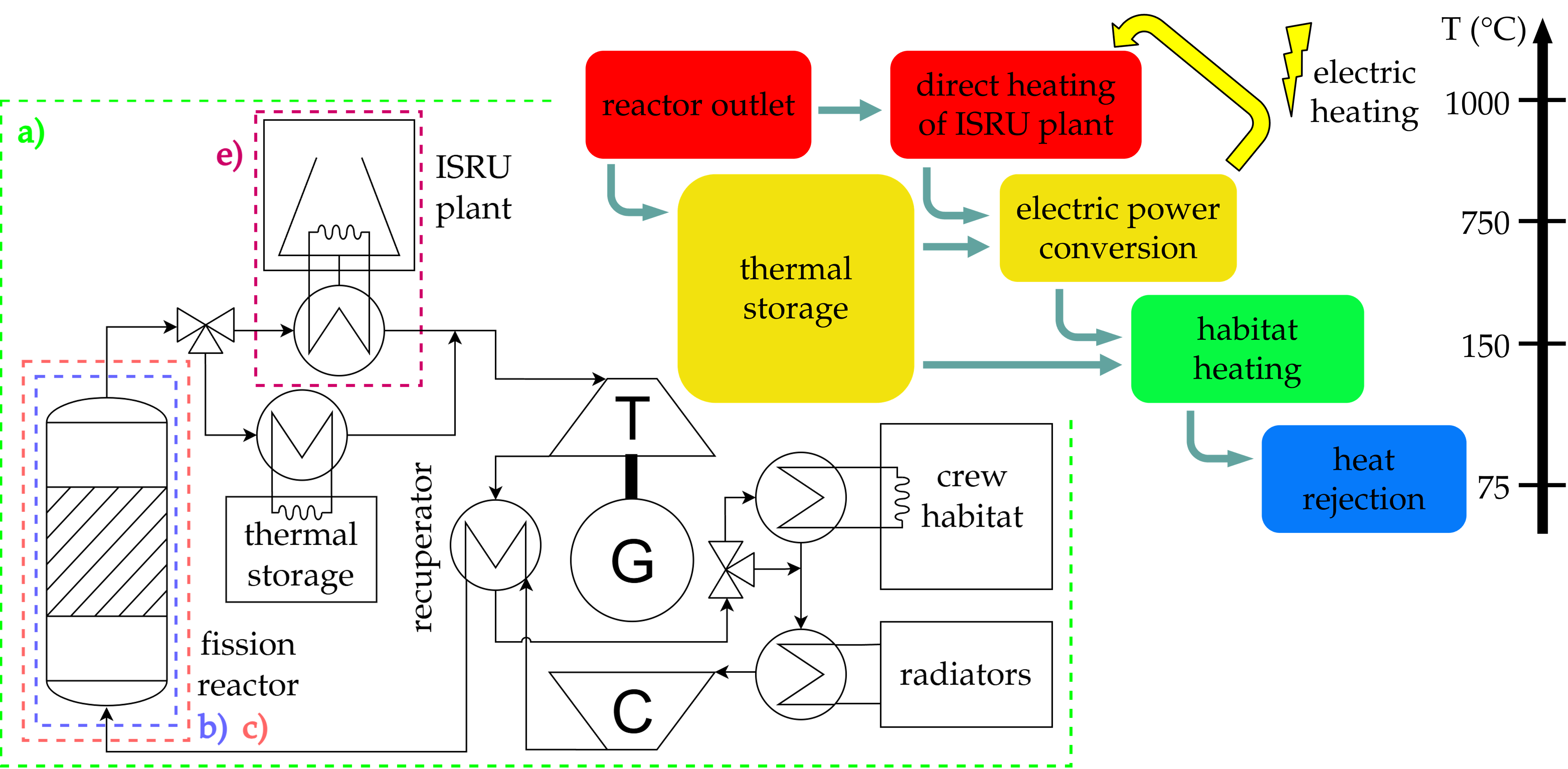"}
		\caption{Schematic of the \ac{mule} co-generation fission power plant concept with adjacent lunar base modules. "T" stands for turbine, "G" for generator and "C" for compressor. Dashed lines indicate the focus of the individual tasks. The flowchart on the top right corner with exemplary temperatures indicates expected temperature levels for each component.}
		\label{fig:mule_overview}
	}
\end{figure}

The goal of the \ac{mule} project is to design a co-generation power plant with a \SI{100}{\kilo\watt} very-high-temperature fission reactor to power a small scale lunar base with a crew of around six astronauts.
Coolant outlet temperatures shall exceed \SI{1000}{\celsius} in order to heat the feedstock of a \ac{mse} \ac{isru} plant to its required operating temperature, which typically lies over \SI{900}{\celsius}.
Since the reactor core needs to be launched on a rocket as an assembled unit, both volume and mass are limiting parameters and need to be minimised.
As a means of additional redundancy during outages, a thermal energy storage can be integrated in the power plant and be heated in periods of low system demand.
The remaining coolant enthalpy can partly be converted into electrical energy inside a gas turbine with a connected generator in a closed Brayton cycle.
Before any waste heat is rejected via radiators, it can be used to heat habitats or other sensitive base infrastructure.
A compressor closes the cycle, while a recuperator increases cycle efficiency and reduces temperature induced stresses for other components.
\autoref{fig:mule_overview} shows an overview of the plant concept.

The very high temperatures constrain the material selection and have a significant impact on neutronics and thus need to be investigated in coupled thermal-hydraulic calculations.
Furthermore, a holistic model of the plant's thermal dynamics is needed to tailor the reactor design.
This necessitates the development of a design workflow and simulation framework for extraterrestrial high-temperature fission power systems.
Five major topics can be derived from this task:
\begin{description}

	\item[a) System Modelling]
		Developing a system model of the co-generation power plant using the \textit{Modelica} language \cite{Modelica2023} to obtain transient responses from all plant components for nominal and off-nominal conditions.
		These include temperatures and pressures throughout the cooling circuits, temperatures and heat fluxes of adjacent components and electrical power generation and consumption.
		Here, the reactor core is modelled on a coarse level and more tailored tools provide more precise results that are fed back into the system code.

	\item[b) Neutronic Reactor Modelling]
		Using the Monte Carlo neutron transport code \textit{Serpent 2} \cite{Leppanen2025}, a high-fidelity model of the reactor core and its structural parts is created.
		It is used for criticality search, burnup, power distribution and also radiation dose calculations.
		The obtained power distribution is used as a volumetric energy source in thermal-hydraulic modelling.
		
	\item[c) Thermal-Hydraulic Reactor Modelling]
		The thermal-hydraulics of the nuclear reactor is analysed using \ac{cfd}.
		Calculated temperatures of the solids are used to check for material limits.
		Outlet temperature and pressure drop of the coolant are passed to the system code.
		The temperature distribution can be passed back to the neutronic modelling to account for feedback effects.
		Transient calculations for scenarios like startup/shutdown and day-night-cycle can be done, which also contribute to a more accurate modelling of the system behaviour.
		
	\item[d) Framework Development]
		The framework couples different codes and models and allow for a more precise simulation.
		For this purpose, \textit{Python} is used, as it is supported on a variety of machines, has a large user base and offers several tools for various applications.
		The framework runs the respective codes, monitors numerical and physical convergence, processes intermediate results and passes them in a suitable format to the other simulation tools, whose boundary conditions need to be updated.
		Additionally, the framework aids design by performing checks on-the-fly on violation of prior set limits and can cancel calculations of designs that will not meet safety or performance criteria.
		For more efficient workflows, post-processing routines are available.
		
	\item[e) Reactor-\ac{isru} Interface]
		To optimise the heating of the \ac{isru} plant, the exact heat exchanger geometry is investigated with \ac{cfd}.
		Of special interest is the phase change of the electrolyte from solid to liquid, when introducing new feedstock.
		
\end{description}

\subsection{Reactor Core Design} \label{ssec:mule_reactor}

\Ac{isru} requires coolant outlet temperatures of around \SI{1000}{\celsius}.
Combined with lunar surface temperatures as low as \SI{-223}{\celsius} during the night \cite{Denevi2009}, Helium is chosen, as it remains gaseous in that temperature range and has a relatively high thermal conductivity. Additionally, it also has a low neutron absorption cross section.
To limit thermal stresses of the reactor core materials, the inlet temperature is set to \SI{850}{\celsius}, leading to a temperature increase of approximately \SI{150}{\celsius}.
In order to support an early lunar base, a desired reactor thermal output of \SI{100}{\kilo\watt_{th}} is assumed, with an anticipated lifetime of at least \num{10} years with minimal maintenance.

These coolant temperatures necessitate a fully ceramic core, with \ac{sic} as the main structural material.
\Ac{triso} fuel with a \ac{uc} fuel kernel is used because of its robustness \cite{Brown2020}, embedded in a \ac{sic} matrix.
The \ac{triso} particles are encapsulated by a hexagonal-shaped \ac{sic} shell with a width of \SI{4}{\centi\metre} and height of \SI{5}{\centi\metre}.
This assembly, which will be referred to as a \textit{fuel compact} in the following, has a circular cooling channel in its centre and is stacked inside a moderator spacer grid with a ceramic compression spring to hold them in place, especially during orbital manoeuvres.
In order to minimise reactor mass for space applications, \ac{heu} with an enrichment of roughly \hbox{$93$ at.-\%} is employed, resulting in a total mass of uranium of \SI{34.31}{\kilo\gram}.

The surrounding radial reflector, moderator spacer grid and axial reflector plugs are made of \ac{beo} due to its high melting point, low density and favourable neutronic behaviour.
The latter share the exact same geometry as the fuel compacts and are stacked beneath and above them.
To control the reactor, six radial and equally spaced control drums are used which consist of \ac{beo} reflector on a \textit{Inconel 718} spline shaft and a \ac{b4c} cylinder sector in a \ac{sic} shell.
Electric motors are connected to the pivoted shaft to actuate the control drums.
The motors are powered by the generator during operation and via a combination of batteries and solar panels during startups.
To allow for thermal expansion, one side of the shaft is mounted on a floating bearing.

\begin{figure}[ht]
	\center{
		\includegraphics[width=0.95\textwidth]{"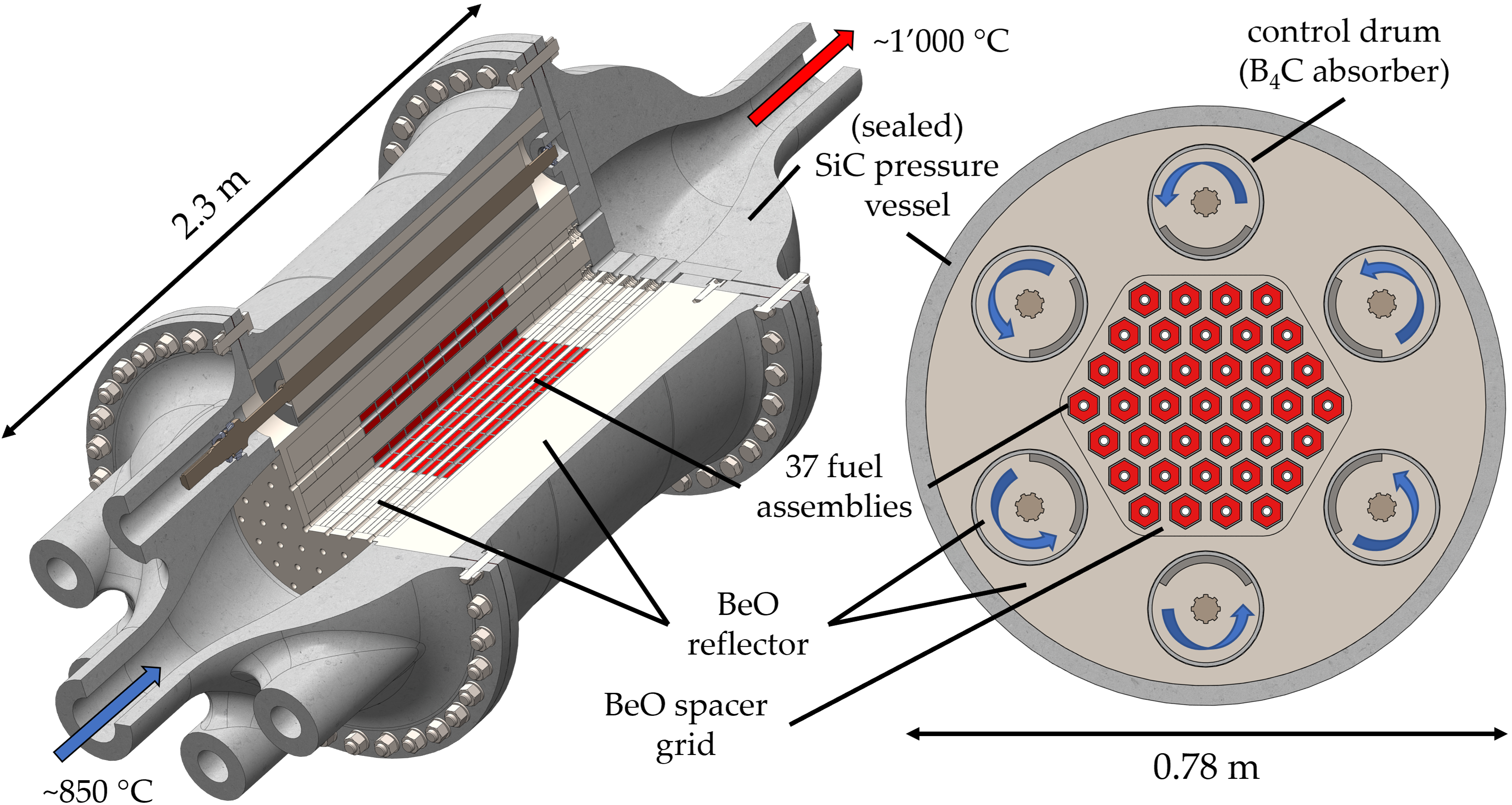"}
		\caption{Sectional cut through length of proposed \ac{mule} reactor design in its shutdown position. The right picture is a horizontal cut near the fuel midplane. Notable parts are annotated and a scale for radial and axial dimensions is given. Electrical motors at the shaft ends and piping flanges are not depicted.}
		\label{fig:mule_reactor}
	}
\end{figure}


The pressure vessel is made of \ac{sic} and is sealed with vermiculite at all flanges.
Fasteners are composed of 1.4401 steel, since they are only used in weakly thermally loaded parts of the reactor.
A biological shield is omitted for the sake of mass optimisation.
Instead, it is proposed to bury the reactor core in some distance to the habitats and utilise the local regolith as a biological shield in order to limit any additional dose to the crew due to the reactor.
The overall dry mass of the reactor in its current design is approximately \SI{2.1}{\tonne}.
An overview of the reactor is given in \autoref{fig:mule_reactor}.

\FloatBarrier
\subsection{Neutronic Reactor Modelling} \label{ssec:mule_model}

\begin{figure}[ht]
	\center{
		\includegraphics[width=0.95\textwidth]{"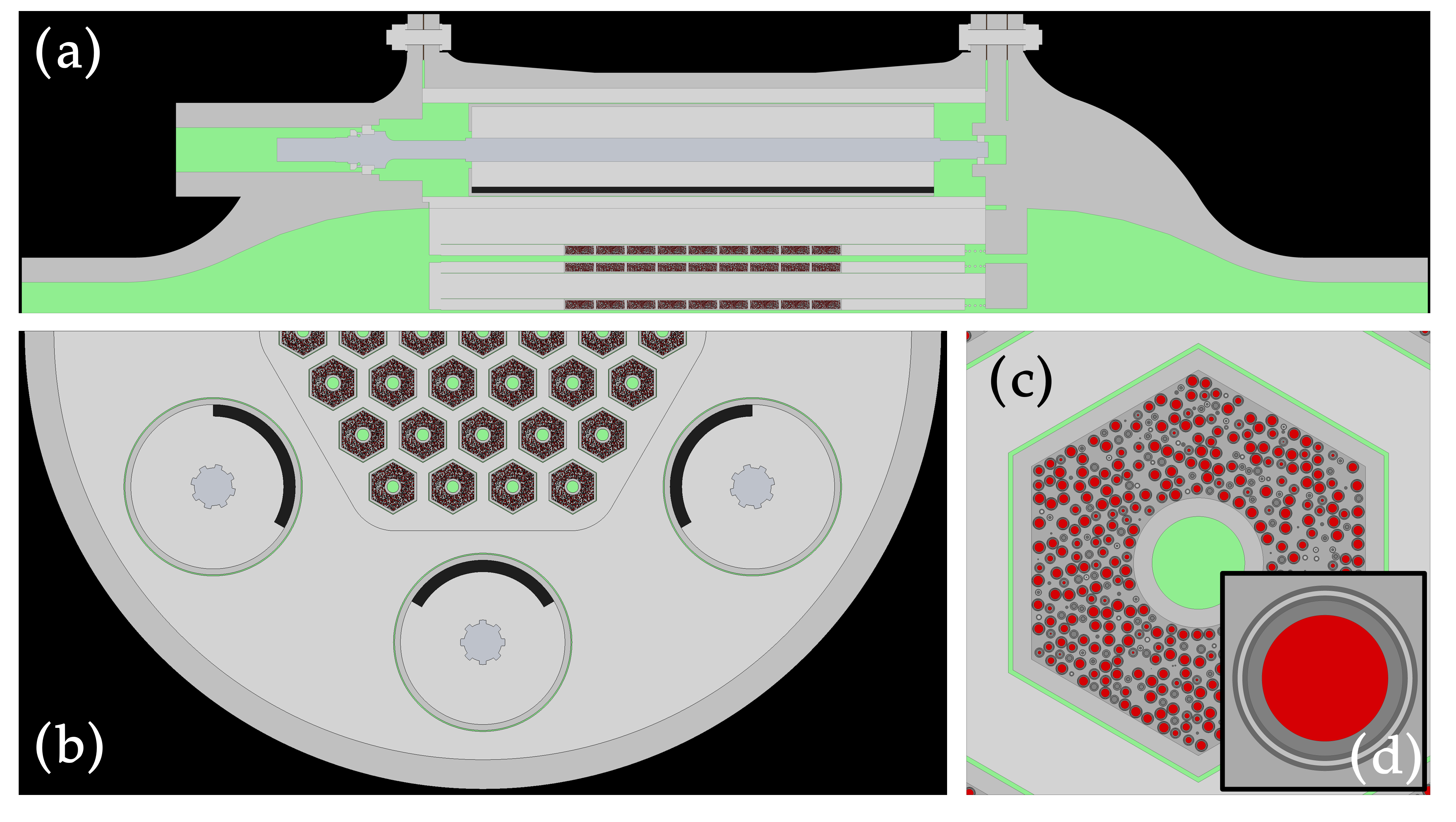"}
		\caption{Neutronic model of the \ac{mule} reactor. (a) is a $y$-$z$ cut through one of the control drums with coolant inlet on the left and outlet on the right hand side. (b) is an $x$-$y$ cut near the fuel midplane with control drums facing inwards ($\varphi_{\mathrm{CD}} = \ang{0}$). (c) shows a fuel compact with its \ac{triso} particles, which are magnified in (d).}
		\label{fig:mule_neutronics}
	}
\end{figure}

In this work, \textit{Serpent-2.2.2} is used as the simulation tool for the neutron transport.
The compactness of the reactor core necessitates a precise geometrical model for \textit{Serpent}, including details like roundings, fasteners, bearings, small gaps, springs etc.
Due to the lack of atmosphere on the Moon, the outside of the reactor is modelled as void.
The \ac{triso} particles inside the fuel compact have been generated with the \texttt{disperse} routine of \textit{Serpent}.
However, as of \textit{Serpent} version 2.2.2, creating dispersed particles in the considered fuel element shape (hex-prism with centred hole) is not available.
Hence, the source code of the routine is modified to include this hex-prism shape with a centred circular exclusion zone.
An overview and detail plots of the neutronic model and the particle distribution are given in \autoref{fig:mule_neutronics}.

\begin{figure}[ht!]
	\center{
		\includegraphics[width=0.8\textwidth]{"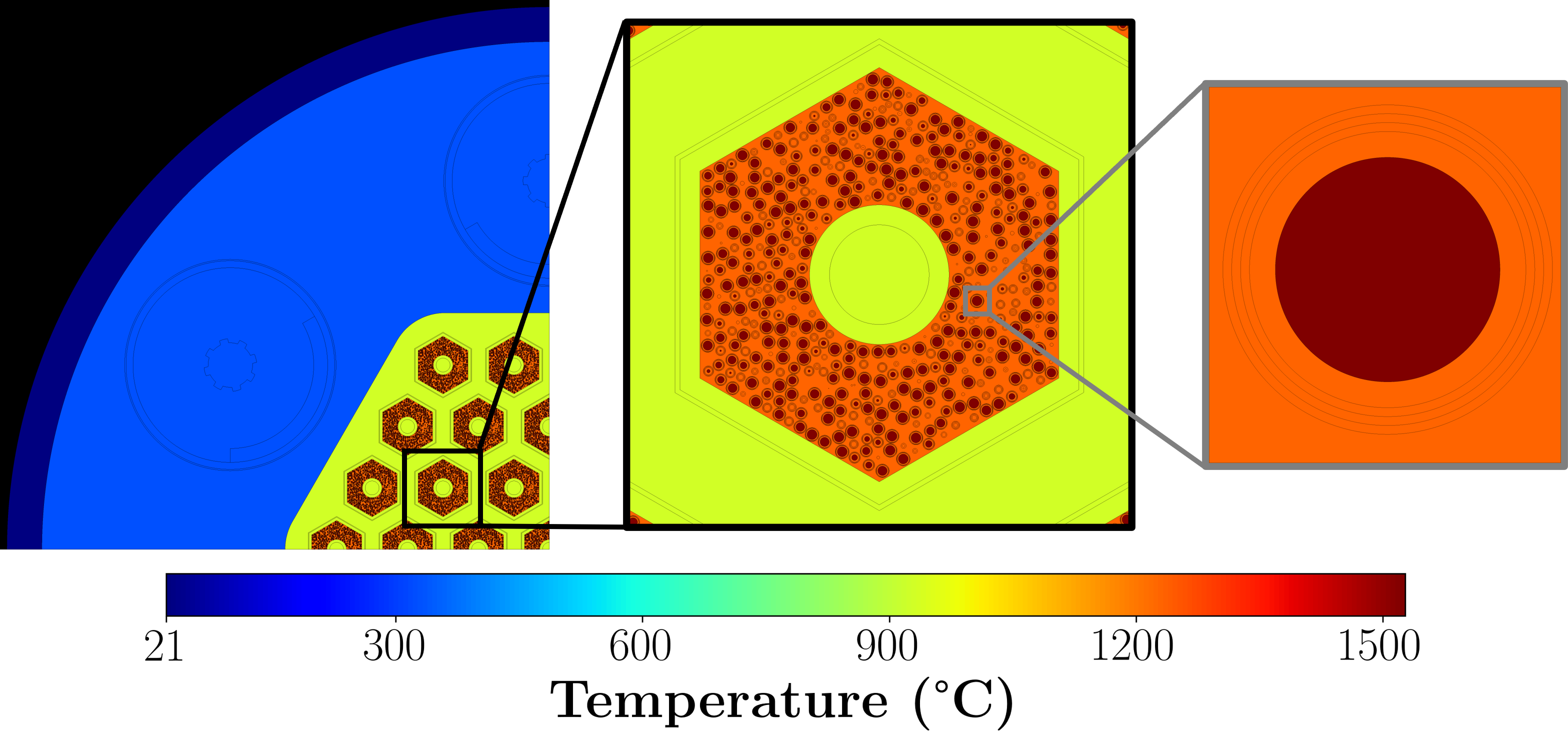"}
		\caption{Assumed temperature distribution in the reactor for the neutronic model.}
		\label{fig:mule_temperature}
	}
\end{figure}

The JEFF-4.0 neutron-induced cross section library has been used, as well as its thermal-neutron scattering law, neutron-induced fission yield and decay data sub-libraries \cite{JEFF2025}.
At this stage of the project, where no coupled thermal-hydraulic calculations have yet been performed, an estimate of the temperatures is used by setting the cross section identifier to the respective temperature for hot reactor conditions.
This sets the fuel kernel to \SI{1527}{\celsius}, the remaining \ac{triso} particle and fuel matrix to \SI{1227}{\celsius} and the fuel compact shell, moderator spacer grid, reflector plugs and coolant to \SI{927}{\celsius}.
The surrounding reflector and control drums are assumed at \SI{327}{\celsius}, whereas the remaining reactor vessel is set to \SI{21}{\celsius}.
Thermal scattering data is applied to graphite in the \ac{triso} particles at \SI{1227}{\celsius}, using the stochastic mixing input option, and to beryllium in \ac{beo} for all affected materials at their respective temperature.
A visualisation of the temperature distribution is given in \autoref{fig:mule_temperature}.
In case of cold conditions, cross sections for \qty{21}{\celsius} are taken.

Steady-state shutdown calculations are done with a control drum angle $\varphi_{\mathrm{CD}}$ of \ang{0}, hence the absorber directly directly facing the fuel zone.
To analyse the lifetime of the design, a burnup calculation with a script-based search for the critical drum position in each burnup step is conducted.
Here, all control drums are rotated by the same $\varphi_{\mathrm{CD}}$, until a \keff\ of \uncpcm{1.0}{5} is reached.
When the critical position is found, the next burnup step is calculated, proceeding from the restart file from the previous step.
The \ac{uc} fuel kernel is set as a burnable material with automated depletion zone division on a fuel compact level.
An initial Monte Carlo volume calculation with 100 billion sampled points provides the volume for all materials in the model, whereas the volume of each subdivided fuel is determined analytically for exact values.
A source from an initial calculation at \ac{bol} in critical configuration is used for faster convergence.
All calculations in the lifetime analysis simulate \num{400000} particles in \num{100} inactive cycles and \num{1500} active cycles with a total power normalisation of \SI{100}{\kilo\watt}.

\FloatBarrier
\section{Results}\label{sec:results}

\begin{figure}[ht]
	\center{
		\includegraphics[width=0.99\textwidth]{"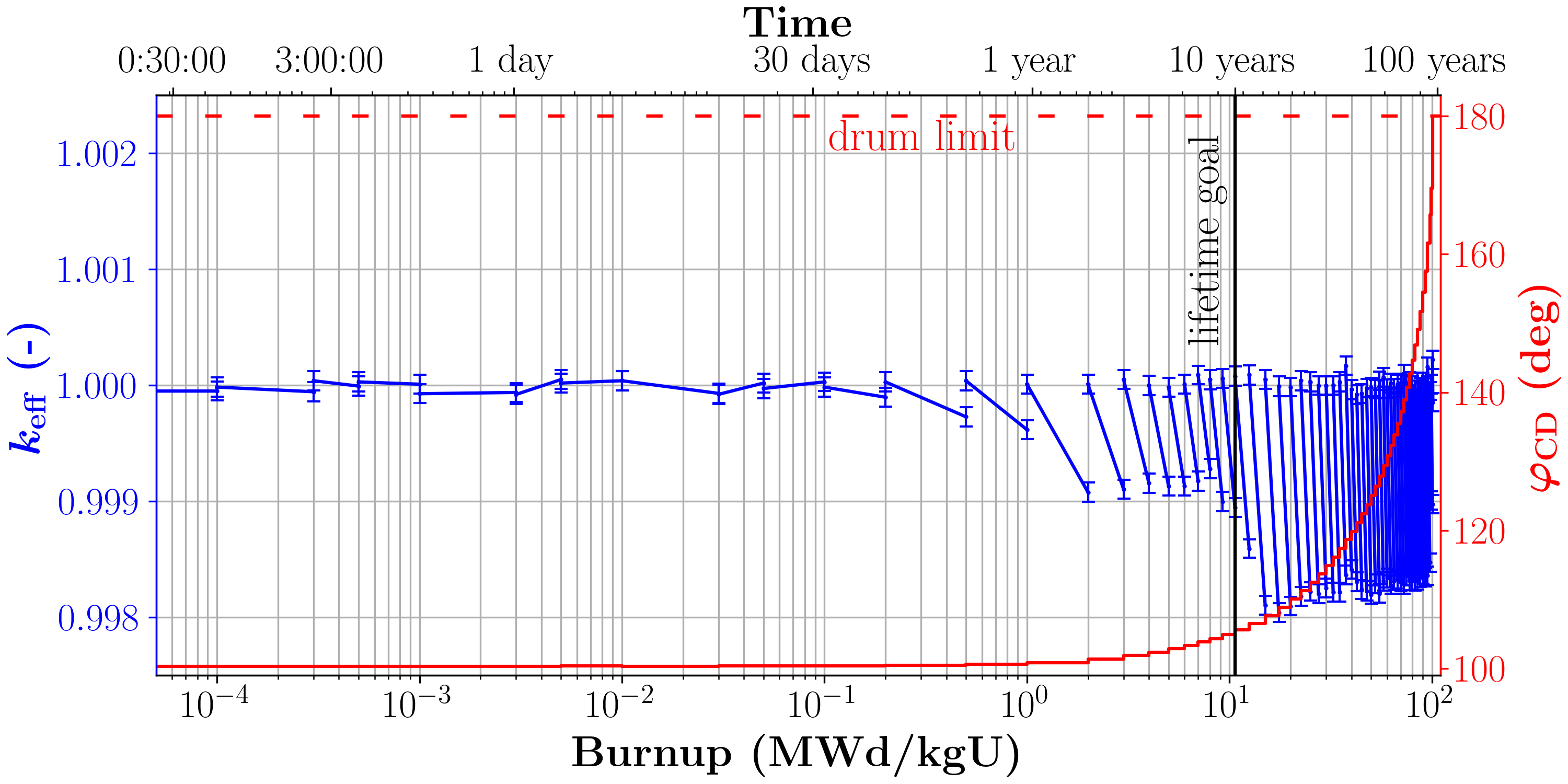"}
		\caption{Results of lifetime analysis with step-wise criticality search. In blue is the obtained \keff\ with $2 \sigma$ absolute uncertainties. In solid red is the calculated control drum angle $\varphi_{\mathrm{CD}}$ for a critical configuration. The horizontal dashed red line denotes the maximum angle of \ang{180} for the control drums, when facing outwards. A black vertical line at 10 years indicates the set lifetime goal.}
		\label{fig:results_s-curve}
	}
\end{figure}

Shutdown calculations ($\varphi_{\mathrm{CD}} = \ang{0}$) for cold conditions at \ac{bol} yield a \keff\ of \uncpcm{0.87751}{4} and for hot conditions \uncpcm{0.87410}{4}, respectively, ensuring shutdown margins.
The following results are obtained in the lifetime analysis with step-wise criticality search.
At \ac{bol}, the critical position is at $\varphi_{\mathrm{CD}} = \ang{100.28}$.
Until \ac{eol} at 10 years (\SI{10.6}{\mega\watt \day \per \kilo\gram U}), changes in the control drum position are minimal, as the critical drum position is calculated at $\varphi_{\mathrm{CD}} = \ang{105.62}$ with \qty{98.41}{\percent} of the initial uranium--235 mass remaining.
As there is still considerable excess reactivity in the core, the calculation is continued until the control drum angle approaches \ang{180} (facing outwards) to estimate the maximal lifetime and further optimisation potential.
The extended \ac{eol} is calculated close to 95 years (\SI{101.0}{\mega\watt \day \per \kilo\gram U}) with \qty{85.00}{\percent} of the initial uranium--235 mass remaining.
The evolution of the control drum angle $\varphi_{\mathrm{CD}}$ throughout the lifetime analysis, as well as the obtained \keff\ is depicted in \autoref{fig:results_s-curve}.

\begin{figure}[ht]
	\center{
		\includegraphics[width=0.99\textwidth]{"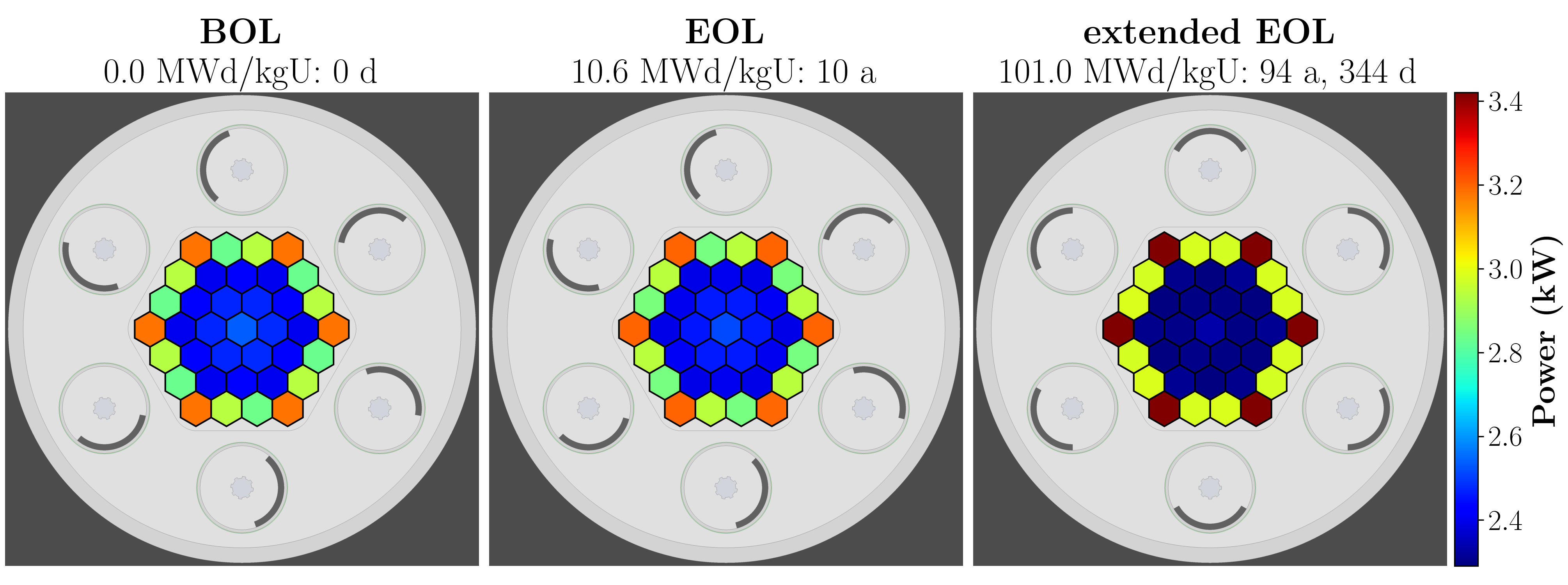"}
		\caption{Power deposition per fuel element for \acf{bol}, \acf{eol} and extended \acs{eol}. The background features the geometrical configuration of the control drums at the respective burnup step.}
		\label{fig:results_power}
	}
\end{figure}

As visualised in \autoref{fig:results_power}, the power distribution peaks in the outermost ring of the fuel zone facing the reflector and in particular in the corner fuel elements with power depositions of up to \qty{3.18}{\kilo\watt} per fuel element.
Due to the small change in $\varphi_{\mathrm{CD}}$ towards \ac{eol}, the power profile does not change significantly.
Hence, the most loaded fuel channel is similarly at \qty{3.20}{\kilo\watt}.
In both cases, the minimum heat load on a channel is \qty{2.40}{\kilo\watt}.
When withdrawing the control drums to their outermost position at the extended \ac{eol}, the thermal flux at the outer fuel ring is increasing, leading to higher power depositions of up to \qty{3.42}{\kilo\watt}.
The distribution is more homogeneous in the outer ring, as the one-sided flux depression due to the absorber pad of the control drum vanishes
The heat load on the fuel elements at the core centre are reduced to values as low as \qty{2.29}{\kilo\watt}.

\FloatBarrier
\section{Conclusion and Outlook}

A co-generation fission power plant concept for an early lunar base with a small crew is presented in this work.
In a first step, a viable design of a ceramic core, gas-cooled very-high-temperature microreactor that can heat an \ac{isru} plant to temperatures over \SI{900}{\celsius} is modelled in \textit{Serpent 2}.
Adaptations to the disperse particle routine of {Serpent 2} allow a realistic distribution of \ac{triso} particles for the proposed fuel element design.
Safe shutdown at \ac{bol} is ensured for cold and hot conditions.
A lifetime analysis in hot conditions with criticality search has shown neutronical feasibility of operational service for nearly 95 years at \SI{100}{\kilo\watt} thermal power output.
Due to the surplus of reactivity at the set \ac{eol}, further mass minimisation can be done in future design iterations.
The obtained power distribution will be used in future thermal-hydraulic \ac{cfd} calculations to estimate a precise temperature distribution in the core for feedback effects.
A system model is expected to mature the plant concept and enable transient analyses of the reactor with its adjacent coolant loop components.

\FloatBarrier
\vfill
\section*{Acknowledgements}

This work has been co-funded by \acs{esa} within \ac{eisi} under contract No. 4000146843. Views and opinions expressed are however those of the author(s) only and do not necessarily reflect those of \acs{esa}, which cannot be held responsible for them.

The authors gratefully acknowledge the support provided by the \href{www.lrz.de}{Leibniz Supercomputing Centre (LRZ)}.

\newpage
\bibliographystyle{ieeetr}
\bibliography{main}

\end{document}